\documentclass[twocolumn]{aa} 
\usepackage{graphicx}
\usepackage{natbib}
\usepackage{latexsym}
\usepackage{amsmath}
\usepackage{amssymb}
\usepackage{amstext}
\usepackage{color}
\usepackage{psfrag}
\def\k{km s$^{-1}$}
\def\ks{km s$^{-1}$~}

\def\s{$^{\prime\prime}$}

\def\cm3{cm$^{-3}$}

\def\3{$^{13}$CO}
\def\2{$^{12}$CO}

\begin{document}

\title{Radio and X-ray study of two multi-shell supernova remnants: Kes~79 and G352.7-0.1}

\author{E. Giacani\inst{1} \footnote[1]{Carrera del Investigador Cient\'\i fico of CONICET, Argentina}
\and M.J.S. Smith\inst{2}
\and G. Dubner\inst{1} \footnotemark[1]
\and N. Loiseau\inst{3}
\and G. Castelletti\inst{1} \footnotemark[1]
\and S. Paron\inst{1} \footnotemark[1]
}

\institute{Instituto de Astronom\'\i a y  F\'\i sica del Espacio
(CONICET-UBA), CC 67, Suc. 28, 1428
     Buenos Aires, Argentina\\
     \email{egiacani@iafe.uba.ar}
     \and
      XMM-Newton Science Operations Centre, ESAC/Selex I. S.,
      Villafranca del Castillo, Spain\\
      \and
       XMM-Newton Science Operations Centre, ESAC/INSA,
      Villafranca del Castillo, Spain}

     \offprints{E.Giacani}

      \date{Received <date>; Accepted <date>}

\abstract{}{}{}{}{}

 \abstract
  {}{We investigate two multi-shell galactic supernova remnants (SNRs),
  Kes 79, and G352.7$-$0.1, to understand the causes of this morphology.}
  {The research was carried out
  based on new and reprocessed archival VLA observations
  and \it {XMM-Newton} \rm archival data.
  The surrounding gas  was investigated based on data extracted from the
  HI Canadian Galactic Plane Survey, the $^{13}$CO Galactic Ring Survey, and the
  HI Southern Galactic Plane Survey.}{The present study  infers that the
  overall morphology of both SNRs is the result of the mass-loss history of
  their respective progenitor stars. Kes 79 is likely to be the product of the
   gravitational collapse of
   a massive O9 star evolving near a molecular cloud and  within the
   precursor's wind-driven bubble,
   while G352.7$-$0.1 should be the result of interactions of the SNR with an
   asymmetric wind from the progenitor together with projection effects.  
    No radio point source
   or  pulsar wind nebula was found to be associated with the X-ray pulsar
   CXOU J185238.6+004020 in Kes 79.
   The X-ray study of G352.7$-$0.1 found
   that most of the thermal X-ray radiation completely
   fills the interior of the remnant and originates  in heated
   ejecta. Characteristic parameters, such as radio flux,
   radio spectral index, age, distance, shock velocity, initial energy,
   and luminosity, were estimated for both SNRs.}
  {}


 \titlerunning {Radio and X-ray study of Kes 79 and G352.7-0.1}

\authorrunning {Giacani et al.}

                 \maketitle
          %
\section{Introduction}

It is expected that between about 70 to 80\% of the Galactic supernova 
remnants (SNRs) originate in the gravitational collapse of massive stars 
that end their lives with powerful supernova explosions of type Ib, Ic, and II. 
In these cases, it is likely that the SNRs initially expand 
inside the wind-driven bubble created by the progenitor stars and 
as a result their evolution differs considerably from that
predicted for an unperturbed surrounding medium. The
presence of forward and reflected shocks as well as peculiar density gradients
 determine the structure, shape, and duration of the successive 
 evolutionary stages of these SNRs. 

Several analytical and numerical studies have been devoted to the exploration
of the interaction between supernova (SN) ejecta and wind-driven
shells (e.g., \citealt{Franco91} and references therein; \citealt{Dwarkadas05}).
 SN 1987A is probably the clearest observational example of the
influence of the past history of the exploding star in
shaping the observed remnant. 

It has been  suggested that the multi-shell appearance observed in the 
SNRs Cygnus Loop and 3C400.2,
for example, can be explained as the product of a SN explosion within
a wind-driven bubble (\citealt{Gvaramadze06} and \citealt{Velazquez01}, 
respectively for the Cygnus Loop and 3C400.2 cases). In the case of 
Cas A, \citet{Borkowski96} modeled X-ray emission originating in a SN explosion
within a wind-blown cavity, and \citet{ Reynoso97} discovered 
the cold envelope of this cavity with HI absorption measurements.
The multi-shell morphology can also result from the proper motion of
the precursor massive star that may cause the star to explode far from
the geometric center of its wind bubble, as proposed by \citet{Gvaramadze06}.
Alternative scenarios can be 
the existence of multiple shocks after the encounter of
the blast wave with a density jump in the surrounding medium, or from simple
projection effects either from three-dimensional structures of the
same SNR or from multiple SNRs along the line of sight 
(as proposed by \citealt{Uyaniker02} to explain the morphology of
Cygnus Loop). In any case, the detailed observational 
study of multi-shell SNRs provides insight into evolutionary and
structural theories of SNRs.

To carry out a multiwavelength imaging and spectral study of this
phenomenon, we selected two Galactic SNRs, Kes 79
(G33.6+0.1) and G352.7-0.1, that exhibit 
clear multi-shell radio structures and also have X-ray emission
detected  in their interiors. The present 
research is complemented by the study of the 
surrounding interstellar gas to investigate the presence of
density inhomogeneities in the interstellar medium (ISM) that may have 
affected the expansion of the SNR shocks.

\section{ The selected sources}

{\sl Kes 79:} This is a Galactic  SNR located at a distance of 7.1 kpc 
\citep{Case98}. At 1.5 GHz and 5 GHz, it appears to consist 
of two concentric incomplete shells with several short  
and bright filaments in their interior \citep{Velu91}. Based on {\it
Chandra} data, \citet{Sun04} describe the associated X-ray emission
as  rich in spatial structures, including filaments, three partial 
shells, and other features named by the authors as  the ``loop'' and the
``protrusion''. Most of the X-rays come from 
the central region.  In addition, {\it Chandra} observations  of Kes 79 in the
0.8-8 keV energy range detected the compact X-ray source
CXOU J185238.6+004020, located close to the geometric center of the SNR
 \citep{Seward03}.
Later {\it XMM-Newton} observations in the 0.3-10 keV energy band
detected pulsations from CXOU J185238.6+004020
 with a period of 105 ms \citep{Gott05, Halpern07}. This source
is likely to be the compact stellar remnant formed in the supernova event, 
although no evidence of a surrounding pulsar wind nebula (PWN) was found 
in X-rays.

It was suggested that Kes 79 is 
interacting with a molecular cloud that partially surrounds
the east and southeast borders of the SNR \citep{scoville87, Green92}. 
 \citet{Green92} reported the detection of bright HCO$^{+}$ J=1--0 
emission from the eastern portion of the remnant at radial velocities close to  $+$105 \k.
 Taking into account
 the spatial and kinematical correspondences, the authors suggested that
 this emission emanates from material associated with the adjacent molecular
cloud that has been shocked by the SNR.
 Further evidence of this interaction
was provided by the observation of a broad and faint absorption feature 
seen in the OH spectrum between $+$95 and $+$115 \ks \citep{green89}.

{\sl G352.7$-$0.1:} This SNR was classified as shell-like type,
 with a size of 8$^{\prime} \times 6^{ \prime}$ and a global spectral index
 of  $\alpha \sim -0.6$ (S $\propto~ \nu ^{\alpha}$) \citep{Green09}.
 A VLA image of G352.7$-$0.1 obtained at 1.4 GHz with an angular
resolution of 34$^{\prime\prime}$ \citep{Dubner93} showed  
 the presence of two concentric ring structures and a conspicuous 
 unresolved bright spot over the eastern limb, whose origin was unclear. 
This bright source was later resolved by 
higher angular resolution observations and 
catalogued in the ``New Catalog of
Compact 6 cm Sources in the Galactic Plane'' \citep{White05} as two
separate point-like sources,  WBH2005 352.775$-$0.153
and WBH2005 352.772$-$0.149. However, their connection with the SNR has yet 
to be clearly established.

\citet{Kinugasa98} presented an ASCA X-ray image of G352.7$-$0.1 
describing the emission as  a shell that roughly coincides with the inner radio shell.
The X-ray spectrum shows prominent K-shell lines from highly ionized Si, S, and 
Ar, interpreted as emission originating in shock-heated optically 
thin hot plasma. \citet{Kinugasa98} proposed that G352.7$-$0.1 is a 
middle-age (2200 years old) SNR located at 8.5 kpc,
 evolving within a pre-existing cavity created by the stellar wind of 
the progenitor.

In what follows, we present new and reprocessed images in radio and in
X-rays of these two SNRs, in an attempt to understand the origin of
the observed characteristics.

\section{ New radio and X-ray data}

\subsection{Radio data}

Kes 79 was observed as  part of a high resolution study
of the neighboring SNR W44, at 74 and 324 MHz using the 
VLA\footnote{The Very Large Array of the National Radio Astronomy
Observatory is a facility of the National Science Foundation operated
under cooperative agreement by Associated Universities, Inc.}  in the 
B configuration on June 15, 2002 and in the A configuration in two 
sessions on August 31 and September 1, 2003. Additional observations 
at 324 MHz were performed in the C and D arrays on December 14, 2002 and 
February 27, 2003, respectively. 
Details of the observations and reduction procedures are described in
\citet{Cas07}. Since Kes 79 is  far  from the phase 
center of these observations, the images at both frequencies were 
corrected for the primary beam attenuation. The synthesized beam, 
rms noise level and other
observational parameters are listed in Table~\ref{table:obs_radio}. 
 It is important to remark that the images at both frequencies are sensitive
to all spatial scales because the size of Kes 79
 ($\sim 10^{\prime}$) is smaller than
the largest well imaged structures at the respective frequencies 
($\sim$36$^{\prime}$ at 74 MHz and $\sim$70$^{\prime}$ at 324 MHz). 

We also reprocessed VLA archival data obtained 
in the direction of Kes 79 at 1.5  GHz.  
The observations  were taken on
July 4, 1989 (BnC configuration, project code AV166), and  on December 24, 1989 
(D configuration, project code AV165). The interferometric data
were combined in the \it uv \rm plane with single-dish information
extracted from the Effelsberg 100 m 1.4 GHz Survey \citep{rei90} to
recover structures at all spatial frequencies. 

To investigate in detail the radio emission close to the X-ray pulsar
candidate while searching for a radio pulsar wind nebula, 
we also reprocessed 4.8 GHz   data corresponding 
to observations carried out with the VLA in the
D array in 1984, September 8 (Project code AD131) and in the D and BnC
configurations in 1989, July 3 and 4, respectively (Project code AV165
and AV166, respectively). These data are only useful for the study
of structures smaller than $\sim 5^\prime$, which correspond to the 
largest angular scale that can be
imaged reasonably well with the array at this frequency. 
 Since no single dish data at 4.8 GHz are publicly available for 
this region, the largest scale structures have not been fully
recovered and therefore no attempt to estimate the total flux density
is made at this frequency.

For the SNR G352.7$-$0.1 we reprocessed VLA archival data at 1.4 and 4.8 GHz. The observations
at 1.4 GHz were performed in the CnD array on February 23, 1991 (project AD260).
The image at 4.8 GHz was produced from the
observations carried out in the DnC configuration on November 3, 1985
(project AH206). Observational parameters are also included in Table~\ref{table:obs_radio}.
As in the case of Kes 79, we did not derive the total flux density of
this SNR at 4.8 GHz because the largest angular scale contributions
are missing.

\begin{table*}
\caption{Observational and derived parameters of the radio data}
\renewcommand{\arraystretch}{1.0}                                                                                
\begin{center}
\begin{tabular}{ccccc}
\hline
\hline

Frequency   & Beam & PA&Noise & Flux Density \\
(MHz)  & (~\s$ \times $\s~) &$^\circ$ & (mJy beam$^{-1}$) & (Jy)\\
\hline
&&{\bf Kes 79}&&\\
74&  39.0$\times$36.0&$-47$  & 60&76$\pm$10   \\
324 &13.3$\times$13.2& $-47$  & 5.8 &39$\pm$8  \\
1500 & 18.6$\times$17.0&$-65.12$  & 1.24&11.0$\pm$1.5 \\
4800 & 14.5$\times$13.0&$-84.8$&0.34&$-$  \\
\hline
&&{\bf G352.7-0.1}&&\\
1400&37.0$\times$33.0&18.33  &1&3.1$\pm$0.3   \\
4800&12.0$\times$ 9.0&-41.34  &0.2& $-$   \\
\hline
\end{tabular}
\end{center}
\label{table:obs_radio}
\end{table*}%

\subsection {X-ray data}

For Kes 79, we analyzed two {\it XMM-Newton} observations, 0204970201 
and 0204970301, which 
were performed on October, 17-18 2004 and on October, 23-24 2004, respectively,
with a 30 ks duration each. In both observations, the MOS cameras were 
operated in full frame mode, with a medium filter. The complete extent 
of the SNR was included in the MOS fields of view. As the PN camera 
was operated in small window mode only part of the remnant was detected,
and hence we did not include these data in our analysis. 

In the case of G352.7$-$0.1, 
we analyzed the {\it XMM-Newton} observation 0150220101 which was performed on October 3, 2002.
Both MOS and PN cameras were operated in full frame mode with a medium filter.
The net exposure times were about
25 ks and 20 ks for the MOS and PN cameras, respectively. 

For both remnants the raw event files were obtained from the {\it XMM-Newton}
Science Archive. The data were processed with the Science Analysis 
System (SAS) version 8.0 and the most up-to-date calibration files. 
The astrometry of the resulting
images was confirmed to be accurate to about 5$^{\prime\prime}$.

\section {Results for Kes 79}

\subsection{Radio emission distribution}

\begin{figure*}
      \includegraphics [width=16cm]{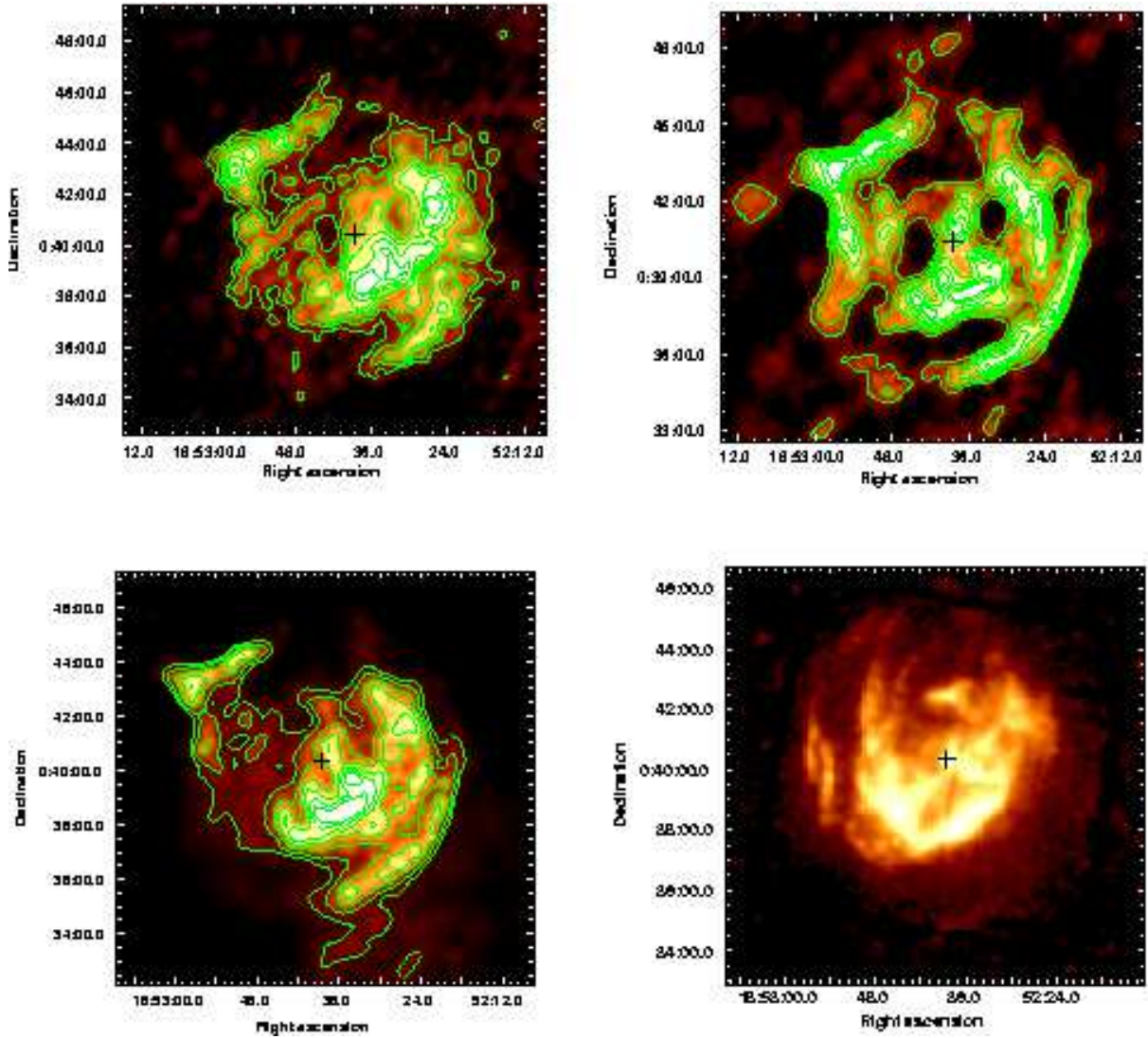}
        \caption{{\it Upper left:} VLA image of Kes 79 at 74 MHz.
 The synthesized beam  is
$39^{\prime\prime} \times 36^{\prime\prime}$ and
the rms noise is 60 mJy~beam$^{-1}$.
Plotted contours go from 0.2 to 0.8 mJy~beam$^{-1}$  in steps of 0.1
mJy~beam$^{-1}$.~{\it Upper right:} The same at 324 MHz.
 The synthesized beam  is $13^{\prime\prime} \times 13^{\prime\prime}$
and the rms noise is 5.8 mJy~beam$^{-1}$.
Plotted contours go from 14 to 50 mJy~beam$^{-1}$, in steps
of 6 mJy~beam$^{-1}$.~{\it Lower left:} The same at 1.5 GHz. Single
dish data from Effelsberg 100 m telescope have been added. The angular
resolution is $19^{\prime\prime} \times 17^{\prime\prime}$ and the rms
noise 1.2 mJy~beam$^{-1}$.
Plotted contours are 10, 12, 16, 22, 28, 34,
40, and 46 mJy ~beam$^{-1}$.  
~{\it Lower right:}
{\it XMM-Newton} image of Kes 79 in the energy range 0.5-5 keV. It has been
square-root scaled to emphasize faint emission.  The plus sign included in all panels
indicates the position of the
point-like X-ray source  CXOU J185238.6+004020.
}
\label{Kes79Radio-X}
\end{figure*}

We present the first radio image of Kes 79  
at 74 MHz  and the highest angular 
resolution image ever obtained at 324 MHz (Fig. 1, {\it upper} panels).
 Figure 1 also includes the reprocessed image at 1.5 GHz ({\it lower left}
 panel), which represents an improvement with respect to
the previously published one  because with the
addition of single-dish data all spatial frequencies are present.
In Fig. 1 ({\it lower right} panel), we have also included for
comparison the {\it XMM-Newton} image obtained in the energy range
 (0.5$-$5) keV.  

The new low-frequency radio images show the  same double
shell structure previously noticed by \citet{Velu91} at higher
frequencies.  The outer radio shell, about 10$^{\prime}$ in
size, is circularly symmetric along the SW and 
W borders of the remnant, while towards the NE and E the 
shell has a protrusion close to Dec $\sim 0^\circ 44^\prime$ that inverts 
the shape into a concave curvature. The 
inner radio shell, about $6^{\prime}.5$ in size, is brighter than the 
outer one at all inspected radio frequencies. It looks incomplete 
and fragmented. Towards the S, this shell consists of a series of 
short arc-like filaments, while towards the E it is fainter and rather 
straight.

The high angular resolution and sensitivity achieved at 324 MHz shows 
  the presence of a small cometary-like feature, about
60$^{\prime\prime} \times 40^{\prime\prime}$ in size, near
the geometric center of the remnant, the
X-ray compact source CXOU J185238.6+004020
lying at the southern border of this structure, whose maximum is  about
$10^{\prime\prime}$ north. We note that the morphology 
of this nebula is  similar to that of some radio pulsar wind nebula
(PWN), motivating us to develop future investigations (see Sect. 4.2).

On the basis of the new images, we estimated
the flux density at 74, 324, and 1500 MHz. The
results are listed in Table~\ref{table:obs_radio} together with 
the observational parameters.  
The errors quoted in the flux density estimates  include the rms noise of each
image and  uncertainties in the determination of the boundaries of the
SNR emission. We note that the total integrated flux
density estimated at 1.5 GHz based on the interferometric observations, 
agrees within the errors 
with earlier estimates obtained from single-dish observations and presented by
\citet{bea69} (S$_{1410 {\rm MHz}} = $ 9 $\pm 3$ Jy) and \citet{alt70} 
(S$_{1414 {\rm MHz}} = $ 13 $\pm$ 0.4 Jy).
This confirms the accuracy of the flux
density estimate and the spectral study discussed in the next section. 

\subsection{Radio spectral index }

We carried out a study of the global radio 
continuum spectrum of Kes 79, based on our low radio frequencies images 
at 74 and 324 MHz and the reprocessed 1.5 GHz data together with those 
taken from the literature. Table~\ref{globalkes79} lists the values 
used to fit the global radio spectrum. We do not include flux
densities obtained from either observations made with instruments 
having insufficient resolution, or those in strong disagreement with 
other measurements made at nearby frequencies.
Most of the measurements for frequencies above 408 MHz  were
brought to the absolute flux density scale of \citet{baa77}. In some cases, an
estimate of the correction factor was unavailable because the
original reference did not list the assumed flux densities of the primary
calibrators. Below 408 MHz, where the systematic error of the scale of
\citet{baa77} is more than 5$\%$, no scaling was
applied. These values have nevertheless been included in our fit because 
they do not exhibit a large spread.

\begin{table*}
\renewcommand{\arraystretch}{1.0}
\centering
\caption{Integrated flux densities on the SNR Kes79}
\label{globalkes79}
\vspace{0.26cm}
{\begin{tabular}[width{10cm}]
{lcllcl} \hline\hline
Frequency  & Scaled flux  &  References & Frequency & Scaled flux & References\\
(MHz)&density (Jy) & & (MHz) & density (Jy) & \\ \hline
30.9\dotfill & 94$\pm$19$^{\mathrm{(a)}}$ & \citet{kas89b}
& 408\dotfill & 44.0 $\pm$ 7.0 & \citet{kes68}\\
74\dotfill &76 $\pm$ 10 $^{\mathrm{(b)}}$ & This work
& 430\dotfill & 34 $\pm$ 10$^{\mathrm{(c)}}$ & \citet{kun67}\\
80\dotfill &101 $\pm$ 23$^{\mathrm{(a)}}$ & \citet{dic73}
& 1410\dotfill & 9.0 $\pm$ 3.0 & \citet{bea69}\\
80\dotfill & 54 $\pm$ 8$^{\mathrm{(a)}}$ & \citet{sle77}
&  1414\dotfill & 13.0 $\pm$ 0.4 & \citet{alt70}\\
80\dotfill & 70 $\pm$ 20$^{\mathrm{(a)}}$ & \citet{dul72}
&  1415\dotfill & 20.0 $\pm$ 6.0 & \citet{cas81}\\
83\dotfill & 76 $\pm$ 12$^{\mathrm{(a)}}$ & \citet{kov94}
&  1500\dotfill & 11.5 $\pm$ 1.5$^{\mathrm{(b)}}$ & This work\\
111\dotfill& 64 $\pm$ 12$^{\mathrm{(a)}}$ & \citet{kov94}
&   2650\dotfill & 8.0 $\pm$ 2.0 & \citet{bea69}\\
160\dotfill& 57 $\pm$ 17$^{\mathrm{(a)}}$ & \citet{sle77}
&  2695\dotfill & 9.0 $\pm$ 0.4 & \citet{alt70}\\
324\dotfill& 39 $\pm$ 8$^{\mathrm{(b)}}$ & This work
& 5000\dotfill & 5.2 $\pm$ 0.4 & \citet{alt70}\\
330\dotfill & 35 $\pm$ 7$^{\mathrm{(a)}}$ & \citet{kas92}
&  5000\dotfill & 7.8 $\pm$ 0.8 & \citet{cas75}\\
408\dotfill & 34.4 $\pm$ 3.4 & \citet{cas75}
& 10600\dotfill & 6.7 $\pm$ 1.5 & \citet{bec75}\\
\hline
\end{tabular}}
\begin{list}{}{}
\item[$^{\mathrm{(a)}}$] No correction to \citet{baa77} scale was applied.
\item[$^{\mathrm{(b)}}$] Flux density scale from VLA Calibrator Manual,
http:/www.aoc.nrao.edu/$\sim$gtaylor/calib.html.
\item[$^{\mathrm{(c)}}$] The correction factor was not available.
\end{list}
\end{table*}

\begin{figure}
\includegraphics[width=7cm]{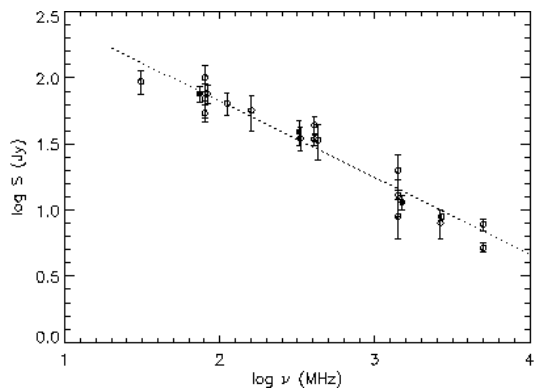}
\caption{Radio spectrum of the SNR Kes~79 obtained from the flux
density values listed in Table~\ref{globalkes79}. The filled circles
correspond to data from the new VLA measurements at 74 and 324 MHz, and the
reprocessed image at 1500 MHz, while the rest of the values were 
taken from the literature and homogenized. The dashed line represents 
the best fit linear function 
to the data points, which yields a
global spectrum for the remnant $\alpha$=$-$0.58$\pm$0.03.
\label{spectrum}}
\end{figure}

The radio spectrum of SNR Kes~79 traced from all frequencies listed in Table~\ref{globalkes79}
follows a single power law $S_{\nu} \propto \nu^{\alpha}$ across
four orders of magnitude in frequency, with an index
$\alpha$= $-$0.58$\pm$0.03, and the fitting is shown in Fig.~\ref{spectrum}.

We also searched for variations in the spectral index across the remnant 
by comparing
the brightness distribution at the different observed frequencies. 
To study the spectral variations as a function of position
within  the remnant, we produced multifrequency images of identical beam
size and shape. For our study, the \it uv \rm data at 324 MHz were tapered to
match the synthesized beam of the data at 74 and 1500 MHz. The images were
also aligned and interpolated to identical projections
(e.g., field center, pixel separation). 

\begin{figure}
\includegraphics[width=7cm]{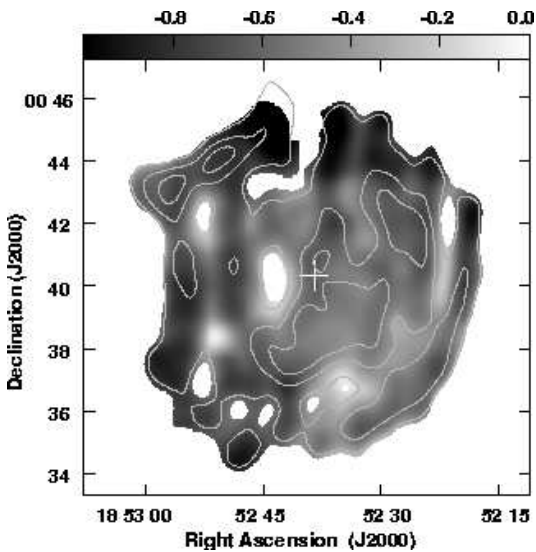}
\caption{Distribution of the spectral index over
Kes~79 between 324 and 1500 MHz, as computed from images made with a
matched restoring beam of 40$\arcsec$.
 For reference,
the 324 MHz contours at 0.1, 0.2, 0.3 and 0.5 Jy~beam$^{-1}$ are plotted.
The position of the CXOU J185238.6+004020 source is indicated by a
white plus sign. Regions with flux density below 4$\sigma$ were clipped.
\label{mapas}}
\end{figure}

Following the same behavior noted above for the global spectral index, 
which is constant over several 
decades in frequency, we find that the spatial distribution of the
radio spectrum  between 74 and 324 MHz, 
is morphologically similar to the spectral map traced between 324 and 
1500 MHz. In Fig.~\ref{mapas}, we show this last comparison degraded to 
an angular resolution of 40$^{\prime\prime}$ to remove unphysical small 
scale fluctuations.  In the comparison, the respective
signals of each radio image were clipped at a 4~$\sigma$ rms noise level.
In Fig.~\ref{mapas}, dark areas correspond to a steep spectrum
 ($\alpha \sim -0.7$), while light gray corresponds to flatter 
indices ($\alpha \sim -0.4$). The
uncertainties vary between 0.04 and 0.1 depending on the intensity of
the  emission compared. A few contours
of the radio intensity at 324 MHz are superimposed for reference. 
The radio spectral index distribution is practically featureless, with
little departure from the global spectral index ($\alpha \sim -0.6$). 
Darker regions observed close to the limb of the SNR are  probably
caused by the larger uncertainties in regions of low emission at 1.5
GHz.

We also conducted a careful study of the radio spectrum close to 
the small nebulosity identified in the image at 324 MHz as a possible
PWN candidate. If this is the nebula created by fresh particles and
magnetic fields injected by the pulsar detected in X-rays (shown with a
plus sign in Fig. \ref{mapas}), a flat
radio spectrum with $\alpha$ between $\sim -0.3$ and 0 is expected for
it \citep{GaenslerSlane06}.  To carry out this search, we used our data at
324 MHz and  the reprocessed archival data at 1.5 and 4.8 GHz. As mentioned in 
Sect. 3.1, at 4.8 GHz the largest scale structures have not been fully 
recovered, but the flux density estimate for this small structure is 
reliable because the size of the studied feature is smaller than the 
largest well imaged structure at 4.8 GHz (5$^{\prime}$).
We obtained a total
flux density for the nebular emission of  0.26$\pm$0.10 Jy, 0.11$\pm$0.03 Jy, and 0.05$\pm$0.06 Jy
at 324, 1500, and 4800 MHz, respectively. The errors quoted include the
rms noise of each image and the uncertainty in the choice of the
integration boundaries. We derive a radio
spectral index $\alpha\simeq -0.6$ between 324 and 4800 MHz. This steep value,
compatible with the global spectral index of Kes 79, rules out
the possibility that this feature is the radio PWN.

In summary, the detailed radio spectral study carried out for the first time
in the SNR Kes79 allows us to conclude that both inner and outer radio
shells have similar spectral properties and no traces of radio
PWN are found near the point X-ray source CXOU J185238.6+004020. 

\subsection{The environs of Kes~79}

We investigated the distribution of the  
interstellar gas in the surroundings of Kes 79. We mapped the neutral gas in 
its vicinity to search for morphological and kinematical signatures of 
atomic material associated with the SNR. The HI data were extracted from 
the Canadian Galactic Plane Survey
(CGPS, see \citealt{taylor2003} for details). The final data have an angular
resolution of about 1$^{\prime}$ and
the velocity resolution is 1.3  km~s$^{-1}$ with a channel separation
of 0.82  km~s$^{-1}$.

\begin{figure}
      \centering
\vspace{1cm}
      \includegraphics [width=7cm]{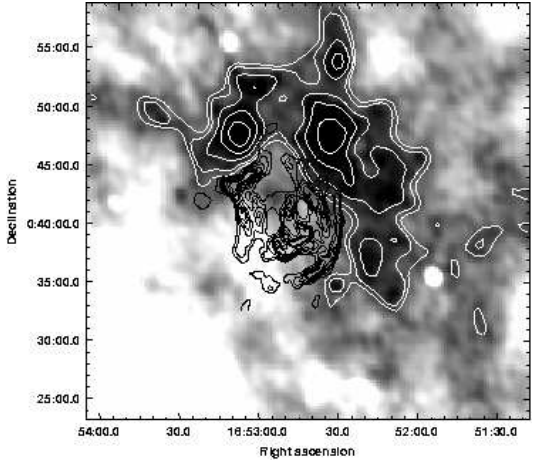}
\caption{Grayscale and contours image of the HI distribution
around Kes 79 integrated between 90 and 99 km~s$^{-1}$. The black contours represent the 324 MHz
radio continuum emission. Grays vary between 80 and 115 K and the plotted 
contours correspond to 103, 106, 109, 113, and 118 K.}
\label{kes79-hi}
\end{figure}

After  careful inspection of the entire HI cube, we find that the only 
channels with an indication of possible physical association are limited to the 
velocity range between +90 and +99  km~s$^{-1}$ (all velocities are 
referred to the local standard 
of rest). Figure~\ref{kes79-hi} shows the HI 
distribution averaged over the aforementioned velocity interval, with some 
contours of Kes 79 at 324 MHz superimposed for reference. In this figure, 
an open HI shell with a radius of about 8$^{\prime}.5$, centered close to 
$18^{\rm h}52^{\rm m}36.6^{\rm s}$, $00^{\circ}41^{\prime}02^{\prime\prime}$
(J2000.0), which embraces most of the periphery of Kes 79, can be
identified. Interestingly, part of this shell exactly matches
the shape of the radio continuum emission along the NE flattened
border, close to $18^{\rm h}53^{\rm m}00^{\rm s}$,
$00^{\circ}45^{\prime}00^{\prime\prime}$.

If we assume that the HI gas in this velocity range is physically associated
with Kes~79, we can adopt 
v$\sim +95$ km~s$^{-1}$ as the systemic velocity of Kes 79. 
By applying the flat Galactic rotation curve of \citet{Fich89} 
(with R$_{\odot}$=8.5 kpc
and V$_{\odot}$ = 220  km~s$^{-1}$), we obtain kinematic distances of
either $\sim$ 6.5 or $\sim$ 7.5 kpc. The difference between these two  values 
is caused by the distance ambiguity in the inner part of the Galaxy and both 
are in good agreement with the distance of 7.1 kpc proposed by
\citet{Case98}. In the absence of any additional discriminator between the
two kinematical distances, in what follows we adopt a distance to 
Kes 79 of 7 kpc.

We can derive a number of parameters characterizing the HI shell. 
By assuming optically thin gas and integrating  the column density 
between +90 and +99 km~s$^{-1}$, the total mass 
in the shell is about 3000 M$_{\odot}$.  A density
of $\sim 8$ cm$^{-3}$ is calculated for the HI shell by assuming that its 
radius and thickness are about 17 pc and 10 pc, respectively (8$^{\prime}$.5
and 5$^{\prime}$ at the adopted distance of 7 kpc).
If this shell was formed by ISM swept up by the SNR shock, then 
the kinetic energy injected into the ISM is about
$1.5 \times 10^{48}$ erg. This value was calculated by assuming an 
expansion velocity of $\sim$ 7 km~s$^{-1}$, as suggested from the HI data.
 This expansion velocity is a lower limit, since confusion from
unrelated foreground and background
emission impedes the detection of possible caps of the expanding shell.
 The estimated values have uncertainties of up to 40$\%$ caused mainly by  
the error in the distance, and secondarily by the choice
of the background level, the expansion velocity, and  the integration 
boundaries. The low expansion velocity found for this HI shell implies 
that this must be an old structure, probably created along thousand of years 
by the action of the precursor's stellar wind. 
 
\begin{figure*}
\includegraphics[width=12cm]{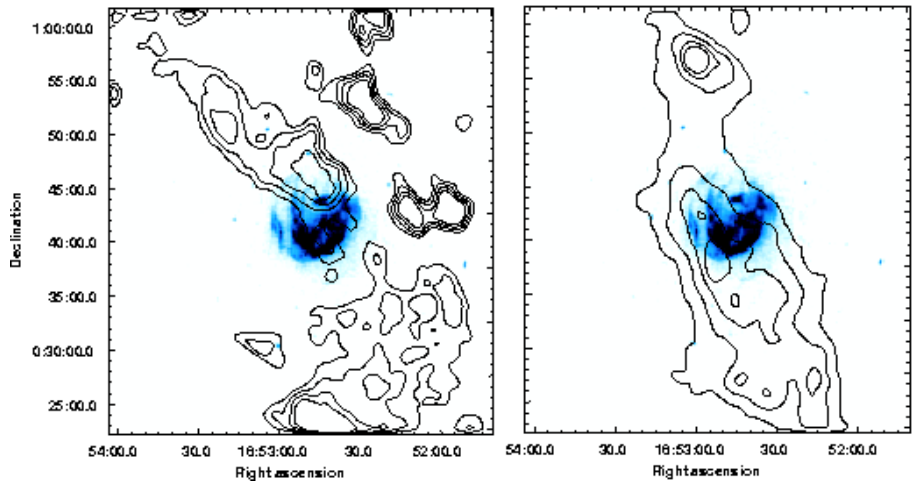}
\caption{{\it Left:}  
 Contours image of the \3 J=1--0 line integrated between $+88.0$ and
$+94.5$ km~s$^{-1}$ with the {\it XMM-Newton} X-ray emission.
 The contours plotted are 1.9, 2.3, 2.8, and 3.3 K \k. {\it Right:} Contours image of the \3 J=1--0 line integrated between $+99.0$ and $+109.0$ \ks with the {\it XMM-Newton} X-ray emission.
 The contours plotted are 3.3, 4.0 and 4.7 K \k.}
\label{nubeco}
\end{figure*}

We also investigated the molecular gas in the surroundings of 
Kes 79 using observations of the \3 J=1--0 line at 110 GHz. The data were 
extracted from the Galactic Ring Survey \citep{jackson06} for which 
the angular and spectral
resolution are 46\s~and 0.21 \k, respectively.
After examining the entire data cube, we found morphological
signatures of a possible interaction between the SNR and the surrounding molecular material when
limited to the velocity intervals ($+$88.0 km~s$^{-1}$, $+$94.5 km~s$^{-1}$) and
($+$99.0 km~s$^{-1}$, $+$109.0 km~s$^{-1}$).
Figure~\ref{nubeco} displays the distribution of the \3 J=1--0
emission integrated from  $+88.0$ to $+$94.5 km~s$^{-1}$ ({\it left}) and 
between $+99.0$ and $+$109 km~s$^{-1}$ ({\it right}) superimposed 
for comparison, on the {\it XMM-Newton} X-ray emission.

The first obvious conclusion is that the optically thin \3 emission does 
not follow the same distribution as the HI.
The most conspicuous molecular feature observed between $+88.0$ and
$+94.5$ km~s$^{-1}$ is  an elongated molecular cloud 
towards the NE of Kes 79, whose maximum  coincides with a region of 
faint diffuse X-ray emission,
suggesting that it is responsible for greater X-ray absorption.
Between +99 and +109 \ks, an extended CO cloud  is detected
crossing the remnant from N to S. This molecular cloud coincides with the 
location of earlier detections of HCO$^+$ and $^{12}$CO emission, and
of the faint OH absorption reported by \citet{green89} and \citet{Green92}.
 
Neither of the two molecular clouds show direct 
kinematic evidence of interaction with Kes 79
(e.g., broadenings in the molecular spectra, asymmetries in the
profiles), although the lack of these signatures  
of interactions is not uncommon in molecular clouds that are known 
to be interacting with SN shocks 
(e.g., as in the case of the SNR 3C 391, \citealt{wilner98}).

An estimate of the atomic and molecular gas column density between us
and Kes 79 was calculated from N$_{\rm H}=2{\rm N(H_{2}) + N(HI)}$ by
assuming local thermodynamic equilibrium, a uniform excitation temperature
T$_{ex} \sim 10$ K for the \3 J=1--0 transition along the line of sight,
and a calibration ratio of N(H$_{2}$)/N(\3) $= 5 \times 10^{5}$
\citep{dickman78}. The hydrogen column density obtained, $\sim 2.2 \times 10^{22}$ cm$^{-2}$, is compatible within errors with the absorbing column density
estimated from the spectral fit of {\it Chandra} and {\it XMM-Newton} data
for the entire remnant ($\sim 1.6 \times 10^{22}$ cm$^{-2}$,
\citealt{Sun04} and $\sim 1.5 \times 10^{22}$ cm$^{-2}$, this work).

\subsection{Analysis of the X-ray emission and  comparison with radio}

\begin{figure*}
      \includegraphics [width=11cm]{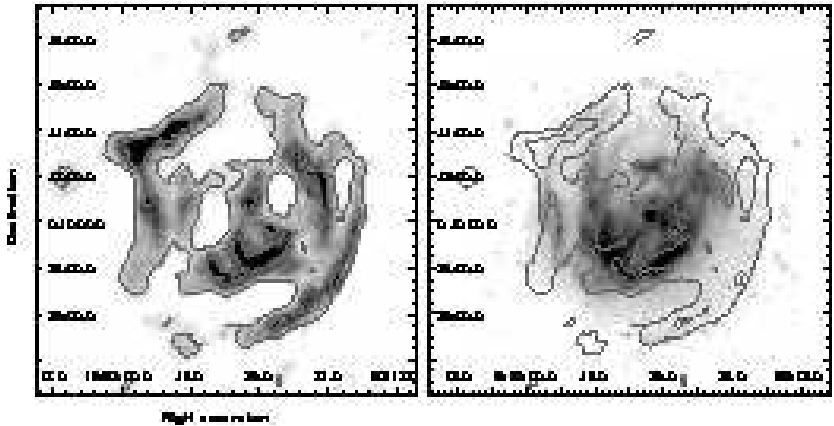}
      \caption{Comparison between the radio emission of Kes 79 at 324 MHz ({\it Left} panel) and the
 {\it XMM Newton} X-ray image obtained between 0.5 and 5 KeV ({\it Right} panel). Some radio contours overlap the X-ray image to facilitate the comparison.}
\label{kes79-comparacion}
\end{figure*}

Figure \ref{kes79-comparacion} shows the detailed comparison between radio 
at 324 MHz and X-ray emission features. It 
can be seen that the most intense and structured  X-ray emission is
confined to a circular region within the perimeter of the inner radio shell,
while fainter and diffuse X-ray emission extends towards the outer
radio limb. As noted before by \citet{Sun04}, an interesting 
correspondence can be
observed between both type of emission, although the agreement is not exact.
The best radio/X-ray matching occurs predominantly along the
southern border of the inner radio shell, where one of the two curved 
bright radio
filaments visible near RA $\sim 18^{\rm h}52^{\rm m}42^{\rm s}$, Dec $\sim
0^\circ38^{\prime}$, has an exact spatial correlation with
bright X-ray emission. The second almost parallel radio
filament in the region, however, does not have an X-ray counterpart.
In the same inner south limb, the X-ray feature called the ``loop'' 
by  \citet{Sun04} (near $18^{\rm h}52^{\rm m}50^{\rm s}$,
$00^{\circ}37^{\prime}$), has a good 
correspondence with a faint radio feature. Towards the E, on the other 
hand, bright twisted X-ray filaments, which were suggested by \citet{Sun04}
to be thin shells viewed edge-on, are almost exactly correlated with
the shape of the radio emission along the outer shell. Another
peculiarity worth noting is that the eastern border of the inner
shell appears to be almost straight in both spectral regimes.
In addition, the flat radio border of Kes 79 near the northern border
of the inner shell, around $ 18^{\rm h}52^{\rm m}38^{\rm s}$,
$ 00^{\circ}42^{\prime}24^{\prime\prime}$,  also 
corresponds with an  X-ray bright feature.

\begin{figure*}
\vspace{1cm}
      \includegraphics [width=14cm]{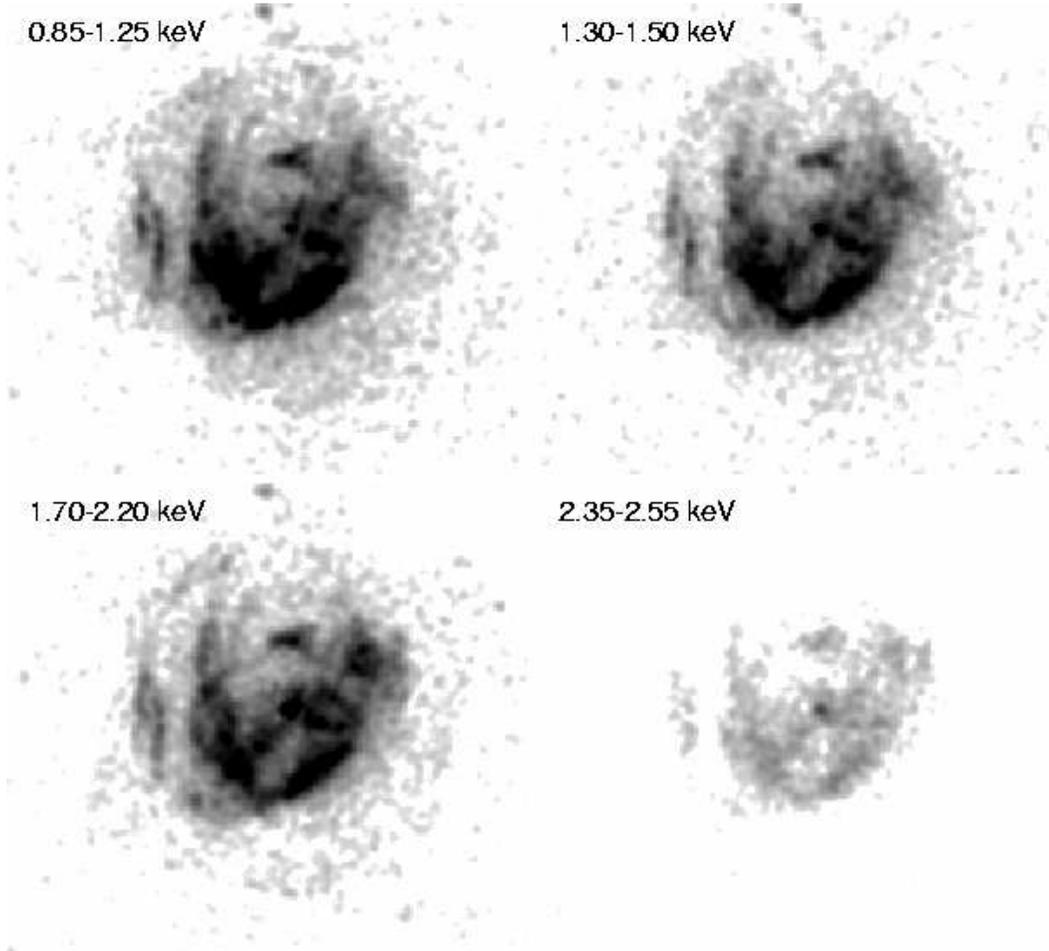}
      \caption{{\it XMM-Newton} EPIC images of Kes 79 in the bands centered at:
       Ne line (0.85$-$1.25 keV), Mg line (1.30$-$1.50 keV),
      Si line (1.70$-$2.20 keV), and S line (2.35$-$2.55 keV).
      The grayscale is the same for all
images to provide a comparative view of the X-ray emission distribution.}
\label{kes79-elementos}
\end{figure*}

Narrow$-$band images of Kes~79 were produced covering the emissions of  
Ne (0.85-1.25 keV), Mg (1.30-1.50 keV), Si (1.70-2.20 keV), and
 S (2.35-2.55 keV) (Fig.~\ref{kes79-elementos}). The same gray scale 
is used in all four images to facilitate the comparison
of data for the respective bands. The central emission is quite similar 
across the four bands and to the broadband emission displayed in 
Figs. \ref{Kes79Radio-X} and \ref{kes79-comparacion}, with the only
exception of sulfur, which is notably weaker than the rest.
From Fig. \ref{kes79-elementos}, it can be seen 
that the Mg  and especially the Si band  emission 
are more prominent in the northwest of the inner shell 
than that of the Ne band, whereas the 
Si emission in the south and southeast is weaker than that of either Ne or Mg.  

\subsection{X-ray spectroscopy}

MOS spectra of both observations (namely 0204970201 and 0204970301)
 were extracted from a circular region centered on 
$18^{\rm h} 52^{\rm m}  41^{\rm s},  +00^{\circ} 40^{\prime} 47^{\prime\prime}$
of $4^{\prime}.74$ radius, encompassing the complete SNR as observed in X-rays.
The background spectra were obtained from four circular source-free regions of $0^{\prime}.73$ radius
located roughly to the north, east, south, and west of the remnant.

We fit the data from both observations simultaneously using a non-equilibrium ionization collisional plasma model \citep{Borkowski01},
VNEI version 2.0, in XPSEC 11.3.
The absorption, temperature, ionization timescale, normalization,
and abundances 
of Ne, Mg, Si, S, Ar, and Fe are free parameters, whereas other elemental 
abundances are frozen to the solar values of \citet{Anders89} after 
verifying that allowing them to vary does not significantly improve the 
model goodness$-$of$-$fit.

\begin{figure}
\vspace{1cm}
\includegraphics [width=8cm]{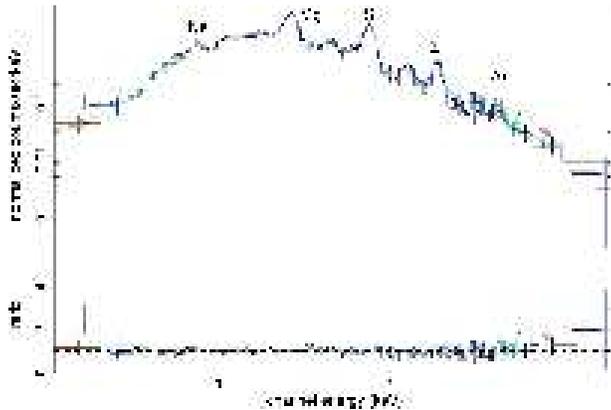}
\caption{EPIC spectra and data-to-model ratios of Kes 79 for observations 0204970201 (MOS1 black, MOS2 red)
and 0204970301 (MOS1 green, MOS2 blue)}
\label{Kes79-espectro}
\end{figure}

We obtained a statistically acceptable fit in the 0.5-5.0 keV band (reduced $\chi^2$ of 1.33 for 991 degrees of freedom)
with parameters values listed in Table~\ref{Kes79xrayfit}; the spectra and data to model ratios are shown in Fig.~\ref{Kes79-espectro}.
The best$-$fit requires lower than solar abundances of Ne, Mg, Si, and
Fe, a slight  overabundance of S and a somewhat larger overabundance of Ar,
although the last value is poorly constrained.
Our results are consistent with those obtained from {\it Chandra} 
data by \citet{Sun04} using the XSPEC VNEI plasma code 
(although the version is different), except for a 30\% higher Ne abundance required for the {\it XMM-Newton} data.

\begin{table}
\renewcommand{\arraystretch}{1.0}
\centering
\vspace{0.26cm}
\caption{The best$-$fit of the global X-ray spectrum of Kes 79. The elemental
abundances listed are those left free parameters in the model.}
\begin{tabular}{lccc}  \hline\hline 
Parameters  & Values  \\
\hline
kT (keV)              & $0.70 \pm 0.01$ \\
$\tau$ (s~cm$^{-3}$)  & $6.2 \pm 0.2 \times 10^{10}$ \\
Ne abundance          & $0.67 \pm 0.02$ \\
Mg abundance          & $0.65 \pm 0.02$ \\
Si abundance          & $0.66 \pm 0.01$ \\
S abundance           & $1.25^{+0.05}_{-0.06}$  \\
Ar abundance          & $5.0^{+0.8}_{-1.6}$ \\
Fe abundance          & $0.54^{+0.01}_{-0.02}$ \\
N$_{\rm H}$ (cm$^{-2}$)   & $1.52^{+0.01}_{-0.02} \times 10^{22}$ \\
\hline
\label{Kes79xrayfit}
\end{tabular}
\end{table}

\subsection{Discussion}

The new high resolution, low frequency radio data confirm the  
multi-shell morphology of Kes 79.
One of the scenarios proposed to explain the presence of
concentric shells in SNRs is that of a SN forward shock propagating
within a wind bubble created by the massive precursor star. The existence 
of an X-ray
pulsar in the interior of Kes 79  confirms that the SN precursor was
a massive star that generated  powerful winds during its lifetime.
As summarized in Sect. 1,  theoretical models predict in these
cases transmitted and reflected shocks that
originated in the interaction of the SN shock with the structural
features left by the stellar wind. In this framework, the outer shell
would represent the blast wave moving into the dense wall of the bubble, while
the inner shell would represent the reflected shock travelling back towards 
the remnant's center. 
 
 Shocks of different origins should have different physical 
conditions (such as for 
example different compression ratios) and this should reflect in the synchrotron
spectrum. From our detailed study, we found that the inner
and outer shells have very similar spectral behavior, with a radio
spectral index close to the global $\alpha\sim -0.6$.
Thus, in terms of the radio spectrum, our study found that the shells are
indistinguishable. 
   
We can investigate the nature of the SN precursor based on the
properties of the observed swept-up HI shell (Fig.~\ref{kes79-hi}). 
For this HI structure, we derived a radius of 17 pc   and a slow
 expansion velocity of 7 \ks. Assuming a linear expansion regime to 
first approximation, we estimate
its age to be about $2\times 10^{6}$ yr. From the
{\it Chandra} X-ray study, \citet{Sun04} estimated that the age of Kes 79 
is between $\sim 3000$ and $\sim 7800$ yr, depending on the
assumed parameters (e.g., shock temperature, shock velocity, density contrast), 
while from radio data, we can set an upper limit to the age by applying
\citet{Chevalier74}'s model, obtaining  t= (R$_{\rm pc}/21.9)^{3.23}
\times 10^{5}$ yr $\simeq 15\times 10^{3}$ yr (where R $\simeq$ 12.2
pc is the assumed radius of the radio remnant). In any case, since 
 the SNR is orders of magnitude younger than the HI shell,  
the detected cold neutral envelope must have formed during the earlier history 
of the star and must be a wind-driven shell.
We can therefore confidently use the physical  parameters of the
associated HI shell to estimate the power of the stellar wind and
hence infer the spectral type of
the progenitor star. Wind blown HI-shells such as the one detected around
Kes 79 have been observed
around other massive OB and WR stars  (e.g., \citealt{Giacani04,
Cappa05, cicho08}).

The kinetic energy of the HI shell structure can be compared
with the mechanical energy released by a
star into the ISM  during the lifetime of the bubble, calculated with
 the equation E$_w =10^{43} \dot M V^2_{\infty}t_{dyn}$/2,
where $\dot{M}$ is the mass-loss rate of the star in $M_{\odot}/{\rm
yr}$, $V_{\infty}$ is the terminal velocity
of the wind in km~s$^{-1}$, and $t_{dyn}$ is the dynamical age of the
structure in years.  If we assume that only one star has powered the 
observed bubble, a single O9 star with a mass-loss rate of
$ \dot M \sim 10^{-7}$ M$_{\odot}$/yr \citep{garmany81} and a wind
velocity of about 2000 km s$^{-1}$ \citep{prinja90} would be sufficient to
sweep up the amount of interstellar gas observed to have accumulated in the
shell, taking into account that the energy conversion efficiency in a 
bubble is no more than 20\%
(see \citealt{Koo92}).
In this case, the mechanical energy released by this star into the
ISM is about 8 $\times 10^{48}$ erg, enough to explain the calculated
kinetic energy of the slowly expanding HI shell (1.5 $\times 10^{48}$
erg).
                                                                                
The elemental abundances derived from the new
X-ray spectroscopy, are in general, consistent with the results of 
\citet{Sun04}, suggesting that we are not observing  shocked stellar ejecta but
swept-up interstellar matter cooling and evaporating in the post-shock
region.    This picture is compatible with a scenario of a middle$-$age
SNR evolving within a wind-driven bubble, where CSM and ISM have been
crossed by forward and reverse shocks.
                                                                                
Based on the measured flux density at 324 MHz, we estimated the
radio luminosity of Kes 79 to be  L$_{\rm R} = 2 \times 10^{34}$ erg
s$^{-1}$, for the assumed distance of 7 kpc and the frequency range
$10^{7}$ to $10^{11}$ Hz. When compared to the
X-ray luminosity derived for this SNR by \citet{Sun04} between 0.5 and 10 keV,
the resulting ratio is over 100. Such high ratios are common in relatively
young SNRs evolving within or at the edges of a molecular cloud, as 
in the case of Kes 79. A large amount of the kinetic energy of a SNR can
transform into bright
emission X-rays, as observed for example in N132D in the LMC \citep{Banas97}
and in the Galactic SNR G349.7+0.2 \citep{Slane02}.                  
                                                                                
\section {Results for G352.7-0.1}

\subsection {Radio emission distribution}

\begin{figure*}
      \vspace{1cm}
            \includegraphics [width=13cm]{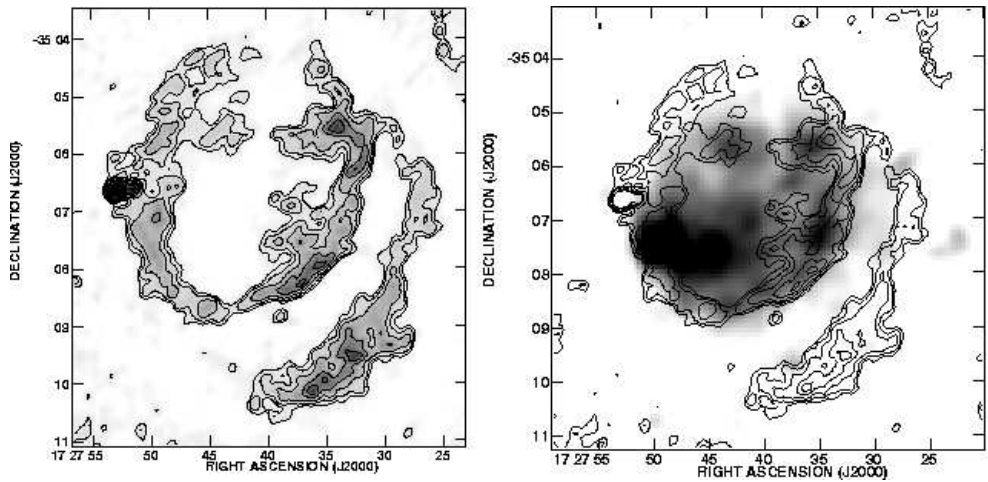}
                  \caption{ {\it Left:} Gray-scale and contour image of G352.7$-$0.1 
at 4.8 GHz. The grayscale ranges from $-$0.1 to 3 mJy beam$^{-1}$ and the contours 
level are: 0.15, 0.3, 0.6, 1.2, and 1.8 mJy beam$^{-1}$. The beam size is 
$12^{\prime\prime} \times 9^{\prime\prime}$ and the noise is 0.2 mJy beam$^{-1}$.
                   {\it Right:} {\it XMM-Newton} image in the 0.15$-$8 keV energy 
band overlaid with radio contours at 4.8 GHz.} 
\label{g352-radio-x}
                  \end{figure*}

The new VLA image of G352.7$-$0.1 at   
4.8 GHz is shown in the {\it left} panel of Fig.~\ref{g352-radio-x}.
 The overall 
appearance of this image resembles that at 1.4 GHz obtained by
 \citet{Dubner93}, but
the higher sensitivity and angular resolution  achieved in the
4.8 GHz image indicate new internal emission, such as the features detected
close to Dec $ \sim -35^{\circ} 05^{\prime} 30^{\prime\prime}$ (which 
strikingly match two X-ray maxima, 
as can be seen in Fig. \ref{g352-radio-x}, {\it right}), as well as
considerable small scale$-$features across the shells. It also 
resolves the
bright spot detected by \citet{Dubner93} at 1.4 GHz at the 
eastern rim of G352.7$-$0.1 into the two point-like sources
catalogued by \citet{White05} as WBH2005 352.775$-$0.153,
centered on $17^{\rm h}27^{\rm m}52^{\rm s}.8$,
$-35^{\circ}06^{\prime}37^{\prime\prime}$, and WBH2005 352.772$-$0.149,
centered on $ 17^{\rm h}27^{\rm m}51^{\rm s}.4$,
$ -35^{\circ}06^{\prime}37^{\prime\prime}.3$.
 As a control of the accuracy of the new
4.8 GHz radio image, we estimated the flux density of these two radio
point sources, obtaining 18 mJy and 3.4 mJy 
for WBH2005 352.775-0.153 and  WBH2005 352.772-0.149, respectively,
values that are in very  good agreement with those of 18.07 mJy and 3.34 mJy 
published by \citet{White05}.  

Because of their peculiar location overlapping the SNR radio shell, it
is important to discern whether one or both of these point-like sources are
part of the SNR structure or they are aligned by chance along the line
of sight. To investigate this issue, we  calculated their radio spectral 
index between 1.4 and 4.8 GHz by carefully matching both images in all
aspects (e.g., {\it uv} coverage, center and pixel alignment).
Since the two radio sources
cannot be resolved in the lower resolution 1.4 GHz image,  
the value of the spectral index is averaged across it. The flux density of 
the spot is 
$\sim$ 40 mJy at 1.4 GHz and $\sim$ 8.8 mJy at 4.8 GHz. In both cases, 
the contribution of the shell emission has been subtracted. The resulting 
spectral index is $\alpha \sim$ $-$1.3. 
To confirm this result, we also computed the spectral index  
using the T-T plot technique 
\citep {Costain60,Turtle62}. This method is useful because it is 
unaffected by absolute calibration and offset variations between the images. 
With this method, the spectral index is $\alpha \sim -1.1$. Although 
the angular resolution
of the available data does not allow us to estimate the spectral index for 
the two point-like sources separately, the obtained value suggests that 
these radio sources are  
extragalactic objects superimposed by chance along the line of sight.
Higher angular resolution observations will help us to discern whether the
peculiar double source appearance can be explained as having originated in
clouds of radio-emitting plasma that were ejected by a distant active
galactic nuclei (AGN) in narrow jets (the type of double$-$lobed
sources called ``DRAGN'', Double Radiosource Associated with Galactic 
Nucleus). For the present study of G352.7$-$0.1, the main conclusion
is that the sources are not related to the SNR.

\subsection{The distance to G352.7$-$0.1}

We investigated the distribution of the interstellar gas in the direction 
of G352.7$-$0.1, with the hope
of finding some morphological concordance between the SNR and the
surrounding matter that would help us to constrain its systemic radial velocity. The HI distribution was inspected by analyzing 
data extracted from the Southern Galactic
Plane Survey obtained with the Australia Telescope Compact Array and
Parkes Radiotelescopes \citep{Mc05}.
The angular resolution of the survey is about 2$^{\prime}$ and the
velocity resolution, 0.82 km s$^{-1}$. The numerous contributions of 
HI emission in this direction
of the Galaxy, close to the Galactic center and above the Galactic
plane, however, impeded the unambiguous identification of associated 
gaseous structures. 
We therefore limited our study to HI absorption. 

HI absorption profiles were traced in the direction of
bright emission radio continuum regions in  G352.7$-$0.1, where we found that   
the most distant HI absorption feature towards G352.7$-$0.1 appears close to 
the radial velocity v$_{\rm{LSR}}$ = $-$90 km~s$^{-1}$. By applying
the Galactic circular rotation model of  
 \citet{Fich89},  this LSR velocity corresponds to the near and far 
distances of $\sim$ 6.8 or $\sim$ 10.1 kpc. We can safely assume    
that 6.8 kpc is the lower limit to the
distance to G352.7 $-$0.1. The upper limit to the distance  can be  
determined from the  Galactic tangent point velocity,  
 v$_{\rm{LSR}}$ $\sim -190$ km s$^{-1}$, which corresponds approximately
to a kinematical distance of 8.4 kpc. Thus,
the distance to G352.7$-$0.1 is 
in between $\sim$ 6.8 and $\sim$ 8.4 kpc.  In a modern 
2-armed spiral pattern of the inner galaxy (e.g., \citealt{Dame08,englmaier08}),
if G352.7$-$0.1 is located at 6.8 kpc, it would lie over the 
``near 3 kpc arm'', while at 8.4 kpc it would be part of the central
bar. Since both of the obtained limits are
plausible, we conclude that a value of $7.5 \pm 0.5$ kpc is an
adequate estimate of the distance to G352.7$-$0.1.

\subsection{Analysis of X-ray emission and comparison with radio}

\begin{figure}
\vspace{1cm}
\includegraphics[width=8.5cm]{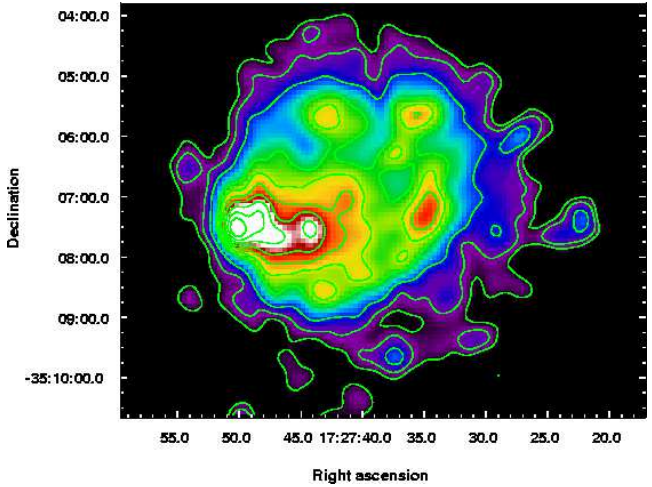}
      \caption{{\it XMM-Newton}
                  EPIC image of G352.7$-$0.1  in the
                  broadband 0.15$-$0.8 keV energy band.}
\label{g352-rayosx}
\end{figure}

Figure~\ref{g352-rayosx} displays the broadband (0.15-0.8 keV) 
{\it XMM-Newton}  EPIC image
of the SNR G352.7$-$0.1. The EPIC, PN and MOS images were merged and 
smoothed with a 8$^{\prime\prime}$
(FWHM) Gaussian. This new high angular resolution image shows
considerable clumpy structures  and diffuse emission filling the interior
of the remnant.  The outermost and faintest X-ray
emission traces an almost circular boundary about 6$^{\prime}$ in
diameter.  Most of the X-ray emission is concentrated in an elongated
feature, whose strongest X-ray peak lies in the eastern border of 
the remnant, centered on about  
17$^{\rm h} 27^{\rm m} 50^{\rm s}.2, -35^{\circ} 07^{\prime} 28^{\prime\prime}.8$.
 Three remarkable bright knots appear
at the north and west border of G352.7$-$0.1 centered on about: 
17$^{\rm h} 27^{\rm m} 43^{\rm s}, -35^{\circ} 05^{\prime} 37^{\prime\prime}$;
 17$^{\rm h} 27^{\rm m} 35^{\rm s}.6, -35^{\circ} 05^{\prime} 38^{\prime\prime}.5$; 
and  17$^{\rm h} 27^{\rm m} 35^{\rm s}.9, -35^{\circ} 07^{\prime} 15^{\prime\prime}.7$.

\begin{figure*}
\vspace{1cm}
      \includegraphics [width=16cm]{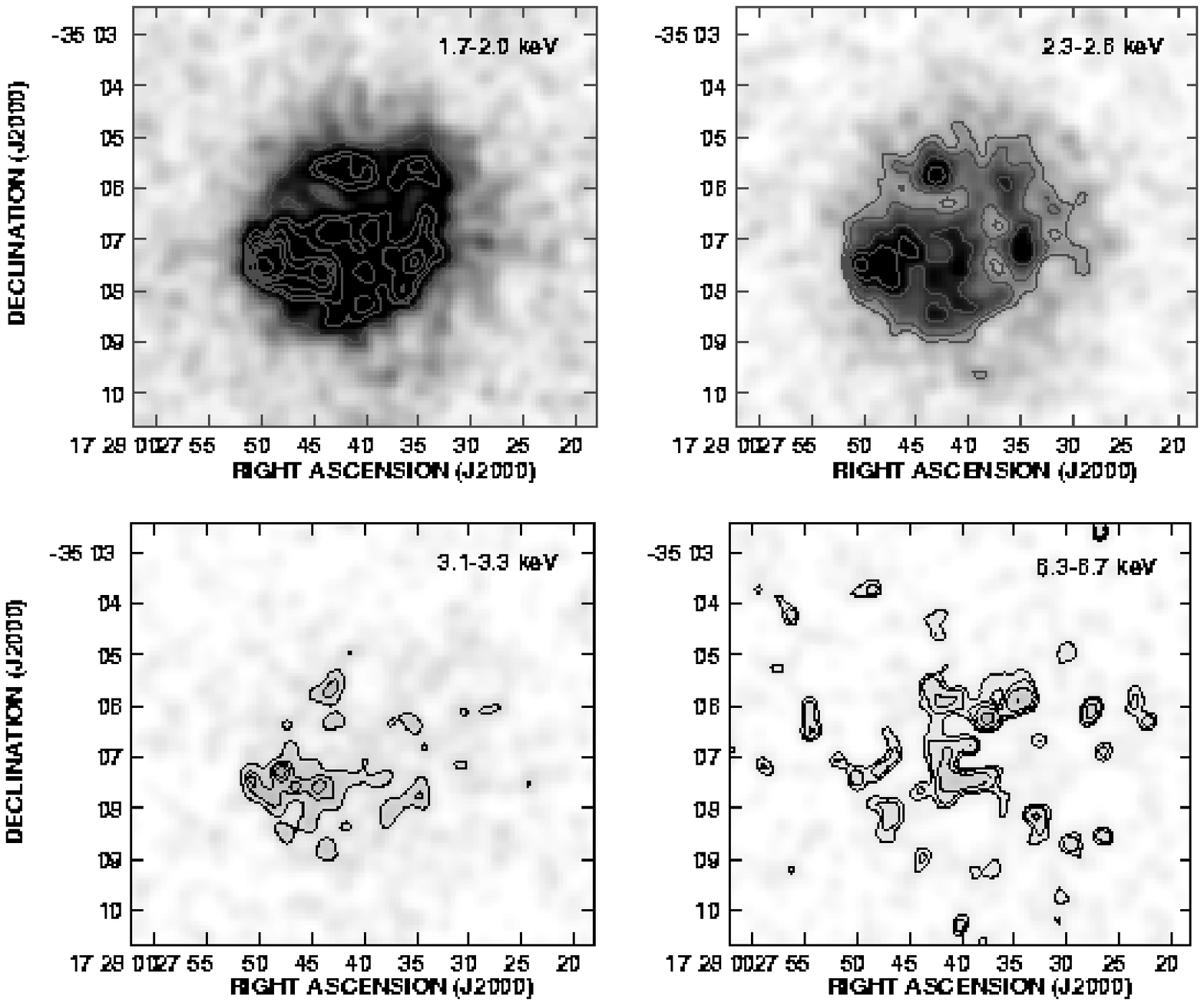}
      \caption{XMM-Newton EPIC images of G352.7$-$0.1 in the bands centered at:
      Si XIII/XV line (1.7$-$2 keV), S XIII/XV line
      (2.3$-$2.6 keV), Ar XVII line (3.1$-$3.3 keV) and 
      Fe K$\alpha$ line (6.3$-$6.7 keV). The grayscale is the same for all
images to provide a comparative view of the X-ray emission distribution.
The contours level varies  from an image to the  other to remark the most
significant features in each one of them.}
\label{g352-elementos}
\end{figure*}

Narrow$-$band images centered on the Si (1.7-2 keV), S (2.3-2.6 keV), Ar
(3.1-3.3 keV), and Fe (6.3-6.7 keV) emission are displayed in Fig.~\ref{g352-elementos}. 
The same grayscale is used for all the 
images to provide a comparative view of the X-ray 
emission distribution. In contrast, the contour levels vary from
an image to the other to emphasize the most significant features in each one. 
 From this figure, can be inferred that soft X-rays  
dominate the emission. The distribution of the emission in the
energy bands centered on the Si and S lines does not show appreciable 
differences 
from the broadband image, while the image centered on the Ar line, instead, 
is remarkable only in the elongated brightest X-ray feature in the southeast. 
The distribution of the
Fe emission line band is more clumpy and anti-correlated with the 
other bands, except for the northern knot  centered on 
 17$^{\rm h} 27^{\rm m} 43^{\rm s}, -35^{\circ} 05^{\prime} 37^{\prime\prime}$, 
which is the only X-ray feature detected in all four X-ray bands. 

Using our high-resolution radio and X-ray images, we are able to
compare for the first time both morphologies in great detail.
Figure~\ref{g352-radio-x} ({\it right}) shows a grayscale {\it
XMM-Newton} image with some radio contours
at 4.8 GHz superimposed. The improved X-ray image shows that X-ray emission is
confined to within the inner radio shell, filling it completely. 
In general, there is no positional correspondence between the radio 
synchrotron and the thermal X-ray emitting plasmas, except for the 
already noted correspondence near Dec. $-35^{\circ}
05^{\prime} 30^{\prime\prime}$, where the bright X-ray
spots coincide with the two radio maxima detected in the interior of
the radio shell, while the
brightest X-ray features, close to 17$^{\rm h} 27^{\rm m} 50^{\rm s},
-35^{\circ} 07^{\prime} 30^{\prime\prime}$, lack any conspicuous
radio counterpart.  The eastern radio point sources, WBH2005 352.775$-$0.153
and  WBH2005 352.772$-$0.149,  do not have
any  X-ray counterpart, confirming that they are unrelated to the SNR.

\subsection{X-ray spectroscopy}

We extracted the MOS1, MOS2, and PN spectra from a circular region of 
radius 2$^{\prime}.5$ centered on
 17$^{\rm h} 27^{\rm m} 36^{\rm s}, -35^{\circ} 06^{\prime} 55^{\prime\prime}$, which covers the entire X-ray emission observed.  
 The background spectrum for the three instruments was taken from an annular
source free region around the remnant with inner and outer radii of 
4$^{\prime}.0$ and 4$^{\prime}.7$, respectively.
 G352.7$-$0.1 is located in the
Galactic ridge region, known to have enhanced thin thermal emission. However, 
as discussed by Kinugasa et al. (1998),
variations in the background with latitude and longitude in this region, can be neglected.

\begin{figure}
      \vspace{1cm}
            \includegraphics [width=8.5cm]{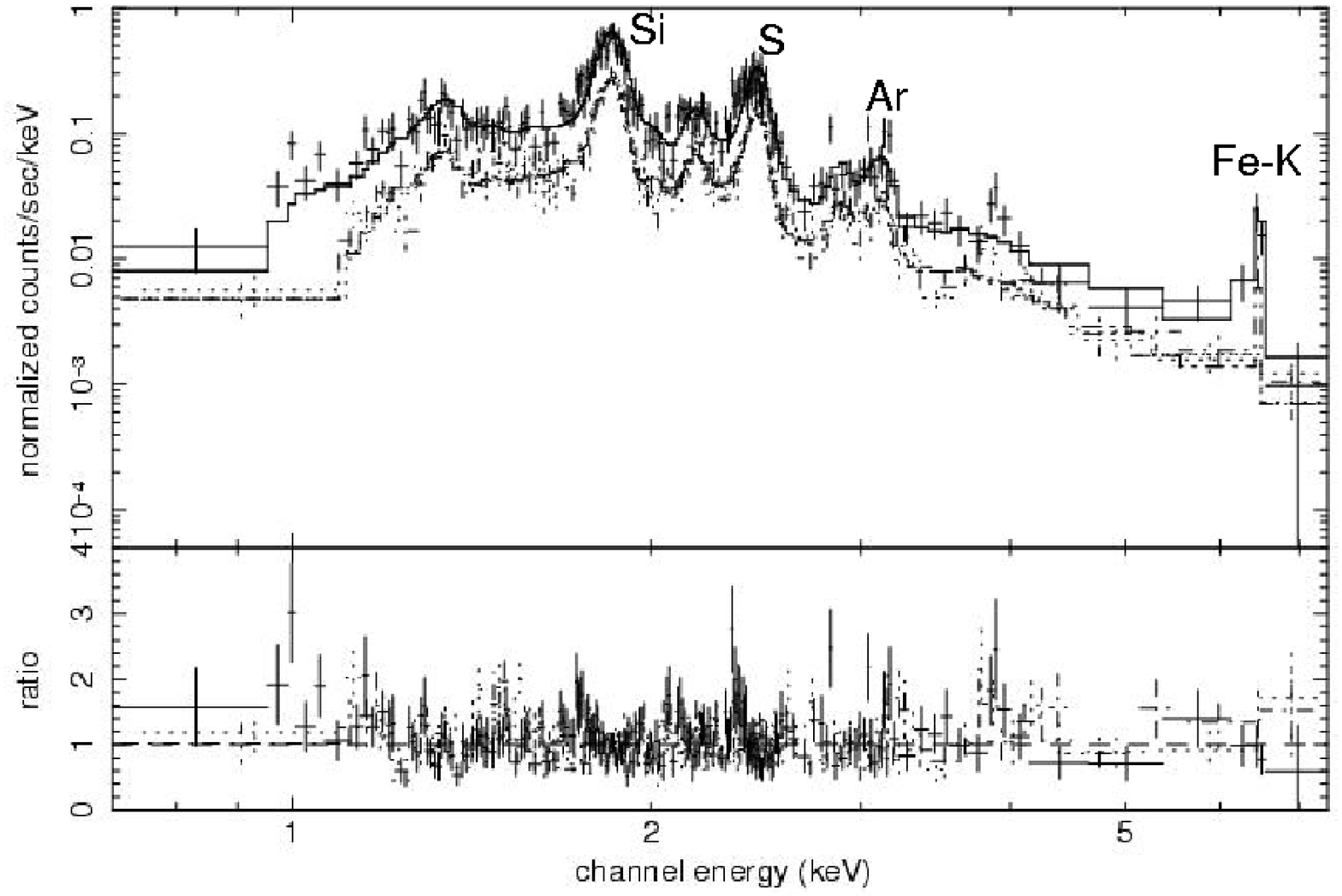}
                  \caption{Energy spectra of G352.7$-$0.1. The lines
for different instruments are: solid for PN, dash-dotted for MOS1, and 
dotted for MOS2.}
\label{g352-espectro}
                  \end{figure}

The data were simultaneously fit in the 0.7-7.5 keV energy band with a 
non-equilibrium ionization collisional plasma model \citep{Borkowski01}, assuming a constant temperature
and single ionization parameter, combined with interstellar absorption.
The absorption, plasma temperature, ionization timescale, normalization, 
and abundances of S, Si, and Ar were considered to be free parameters 
in the model.
We verified that allowing other elemental abundances to vary did not 
significantly improve the model fit, hence there were frozen to the solar 
values of \citet{Anders89}. An additional zero-width Gaussian component
was needed to model the Fe K emission resulting in a line centroid at
6.46 $\pm$ 0.03 keV, which would be consistent with emission caused by
 Fe K$\alpha$ fluorescence. We obtained a statistically good fit
 (with a reduced 
$\chi^2$ of 1.25 for 594 degrees of freedom), the results of which are listed
in Table~\ref{G352xrayfit} and plotted in Fig.~\ref{g352-espectro}. 
As can be seen, the best fit requires 
significant over abundances of S, Si, and Ar with respect to their solar
values, which indicates the presence of ejecta material. The 
 Fe L abundance does not differ significantly from the solar value, which
 argues in favor of a core collapse SN. However, this cannot fully 
constrain the progenitor explosion because  the
relative lack of Fe could only be explained if the innermost ejecta 
layers have not
yet been shocked \citep{Rakowski06}. Regarding the Fe K line at an energy
of 6.46 keV, it has also been observed in
other young SNRs such as Tycho \citep{Hwang97}, Kepler \citep{Kinugasa99}, and
RCW86 \citep{Bamba00}, and also in G344.7-0.1 \citep{Yamauchi05}. The
origin of this line is not well understood. 
  
\begin{table}
\renewcommand{\arraystretch}{1.0}
\centering
\vspace{0.26cm}
\caption{The best$-$fit model of the global X-ray spectrum of G352.7$-$0.1. The 
elemental abundances listed are those considered to be free parameters of 
the model.}

\begin{tabular}{lccc}  \hline \hline
Parameters  & Values  \\
\hline
 kT (keV) & 1.9 $\pm$ 0.2 \\
 $\tau$ (s~cm$^{-3}$)  & 4.5 $\pm 0.5 \times 10^{10}$ \\
 Si abundance  & 2.4 $\pm$ 0.2 \\
 S abundance  & 3.8 $\pm$ 0.3  \\
 Ar abundance & 4.7 $\pm$ 1.2 \\
 N$_{H}$ (cm$^{-2}$) & 2.6 $\pm$ 0.3 $\times$ 10$^{22}$ \\
\hline
\label{G352xrayfit}
\end{tabular}
\end{table}

We searched for spectral variations across the remnant by individually
extracting the spectrum of each of the most conspicuous X-ray
features and the diffuse emission. This procedure was performed
separately for each instrument. We found no significant difference in the fit
parameters for any of them compared with the global model.

Based on the emission measure (EM) determined from our spectral
fitting and a distance of 7.5 kpc,  we can
estimate the electron density of the plasma to be n$_{\rm e}$ =
(EM/V)$^{1/2} \simeq 0.3~ $cm$^{-3}$,
where V is the volume of the X-ray emitting plasma
considered in this case to be a sphere of 2$^{\prime}$.5 of
radius, and  the  hydrogen number density,
n$_{\rm e}$ = n$_{\rm H}$,
 for simplicity. The age of the remnant, estimated to be $\tau$/n$_{\rm
e}$, is found to be 4700 yr, and the mass of the X-ray emitting gas is
M=  n$_{\rm e}$Vm$_{\rm H}$ $\simeq 10 {\rm M}_{\odot}$, where
 m$_{\rm H}$ is the hydrogen atomic mass.
For a shock temperature of kT = 1.9 keV, the shock velocity
would be v$_{\rm s}$ = (16kT/3$\mu$ m$_{\rm H}$)$^{1/2} \simeq 1300$
km~s$^{-1}$ (with a mean atomic weight  $\mu$=0.61).
For the preceding numerical values,
we estimated the supernova explosion energy E to be about
10$^{50}$ erg, which is the standard value for SN explosions.
Based on the NEI fit to the spectrum of the entire remnant
in the (0.7-7.5) keV energy band, we estimate the X-ray luminosity to be
 L$_{\rm X}\sim 8.4\times 10^{34}$ erg s$^{-1}$ for a distance of 7.5
kpc. This value can be compared with the radio luminosity calculated
on the basis of the flux density measurement at 1465
MHz carried out by \citet{Dubner93} to obtain 
L$_{\rm R}= 4.6 \times 10^{33}$ erg s$^{-1}$, and a ratio
L$_{\rm X}$/L$_{\rm R} \sim 20$.

\subsection {Discussion}

\citet{Kinugasa98} described the X-ray emission of G352-7$-$0.1  as  
shell-type  with 
a ring morphology similar  to that observed in the radio band. 
However, our new sensitive {\it XMM-Newton} image of this SNR indicates
 that the X-ray emission completely fills the interior of the radio remnant.
 Although the centroid of the X-ray 
emission lies over the eastern half of the remnant, both the observed 
morphology and the thermal nature of the emission with a flat radial 
temperature profile, allow us to re-classify 
G352.7$-$0.1 as a mixed-morphology remnant, i.e., shell-type in radio and 
filled-center in X-rays \citep{Rho98}. 
The X-ray emitting plasma has not yet reached the ionization equilibrium, as 
expected for young remnants, and has enhanced metal abundances,  indicative
of an ejecta origin.
 
\begin{figure}
     \includegraphics [width=8cm]{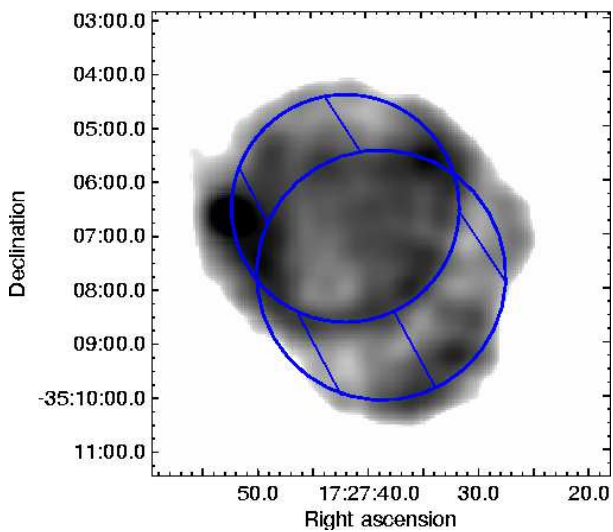}
       \caption{Model of ``barrel-shaped'' SNR overlapping the VLA image 
 of G352.7$-$0.1 at  1.4 GHz from \citet{Dubner93}.}
\label{esquema}
\end{figure}

The observed  morphology matches a ``barrel-shaped'' model fairly well  
for the radio SNR (that is, a structure with cylindrical geometry in space, 
as discussed by \citealt{Manchester87} and \citealt{Gaensler98})
with a peculiar viewing angle as depicted in Fig. \ref{esquema}.  
To produce this figure we  used the VLA image at 1.4 GHz from \citet{Dubner93},
which recovered all the flux density. The radio emission 
appears to be enhanced in the directions where our 
line of sight crosses a longer path through the SNR.
As proposed by \citet{Manchester87}, the bilateral enhancements observed in 
``barrel-shaped'' SNRs can originate 
in a biconical flow from the pre-supernova star, 
from the supernova explosion itself, or from an associated pulsar or X-ray binary system, and is probably formed early in the life of the remnant.
In the case of G352.7$-$0.1, which is known not to harbor a pulsar
in its interior, the origin of the observed appearance must be connected with 
the  mass$-$loss history of the progenitor star. In this context,
 \citet{Manchester87} proposed that a biannular or double ring morphology
 can be produced  when the SN 
shock expands through the axially symmetric wind of 
a red supergigant (RSG) phase of a massive star.
Typical parameters for the RSG phase are a wind velocity of about 
20 km s$^{-1}$  and a
lifetime of about 3$\times 10^{5}$ yr \citep{Smith94}. This wind perturbs a 
region with a radius of 6 pc, a size comparable to the radius of G352.7$-$0.1.
The X-ray spectrum results, which favor a type II explosion
for this SN, confirm a scenario that involves a massive star precursor.
                                                                                
\section{Conclusions}

\begin{table*}
\caption{Main results}
\renewcommand{\arraystretch}{1.0}

\begin{center}
\begin{tabular}{lll}
\hline
\hline

& Kes 79 & G352.7$-$0.1 \\
\hline
Radio flux density &at 74 MHz = 76 Jy  & \\
& at 324 MHz = 39Jy & \\
& at 1.5 GHz = 11 Jy & \\
Radio spectral index& -0.58&\\
Distance & 7 kpc& 7.5 kpc\\
Age& $<$ 15000 yr & 5000 yr\\
X-ray-to-Radio Luminosity& $\sim$ 100& $\sim$ 20\\
Elemental abundances& Shocked CSM and ISM& Ejecta\\
Scenario& Core collapse of an O9 star& Interaction with an asymmetric progenitor wind \\
&within a wind-driven bubble& resulting in a ``barrel-shaped'' SNR\\
& plus projection& plus viewing angle and projection\\
\hline
\end{tabular}
\end{center}

\label{table:conclusions}
\end{table*}

We have analyzed new high resolution and sensitivity VLA images as 
well as archival 
radio and {\it XMM-Newton} X-ray data of two galactic SNRs, Kes 79 and
G352.7$-$0.1, which share a common multi-shell radio morphology. 
In addition,  we have investigated the ISM in the direction of both 
remnants to probe the conditions of their surroundings.

 New radio features detected in 
the image of Kes 79 at 324 MHz considerably improve the correspondence 
with the X-ray emission distribution of this remnant.
Based on our new radio data and 
flux density estimates taken from the literature, we derived a global spectral
index of $\alpha =-0.58\pm 0.03$, constant over four decades in frequency 
with no significant turnover down to 74 MHz. The new 74 and 324 MHz data
and the reprocessed 1500 MHz data were combined to perform the 
first careful, spatially resolved study of the synchrotron spectrum in Kes 79, concluding that the radio spectrum is homogeneous, with no detectable variations 
associated with shells, maxima, or any other morphological feature. We 
have also 
confirmed that down to about 5.8 mJy beam$^{-1}$, neither a radio counterpart
to the X-ray pulsar CXOU J185238.6+004020, nor a surrounding PWN is observed.
Our study  confirms that Kes 79 is the result of a core-collapse SNe 
evolving close to a molecular cloud and within
the wind-blown bubble created by its precursor.
The average X-ray spectral properties are well described by
a non-equilibrium ionization collisional plasma model with constant
temperature across the remnant and solar abundances.
The observed  multishell appearance of Kes 79 can be described as
the final product of possibly a O9 star exploding within the cavity
created by previous episodes of mass loss and evolving near the parent molecular
cloud. The precursor wind has
swept up the surrounding gas forming a thick HI shell. 
A scenario in which the multishell morphology originated 
from multiple shocks produced 
after the encounter of the blast wave with a density jump in the
surroundings, could explain the observations. 

In the case of the SNR G352.7$-$0.1, the new VLA image at 4.8 GHz uncovered  considerable clumpy 
structures on small scales, confirming that the bright spot previously observed 
on the eastern limb is produced by a double extragalactic radio source.
 The new {\it XMM-Newton} image shows 
 several knots of emission and diffuse thermal emission filling
the interior of the remnant. The X-ray spectral investigation inferred 
enhanced abundances, which imply
the presence of SN ejecta in this SNR. From the morphology and spectral 
properties of G352.7$-$0.1, we classify this remnant as 
belonging to the mixed-morphology category. The observed appearance can be
the consequence of the propagation of the SN blast-wave within an axially
symmetric stellar wind blown out by the precursor star, forming  
a ``barrel-shaped'' SNR that is observed at a peculiar viewing angle. 

The study of the surrounding interstellar gas around both SNRs allowed us to refine the distance estimate towards 
these two Galactic SNRs, obtaining a distance of $7 \pm 0.5$ kpc for Kes 79, and $7.5 \pm 0.5$ kpc for G352.7$-$0.1.
Table~\ref{table:conclusions} summarizes the main findings of the present
study.

\begin{acknowledgements}

We thanks the anonymous referee for helpful comments.This research was 
partially funded by Argentina Grants awarded by 
 ANPCYT, CONICET and University of Buenos Aires (UBACYT A023). 
This work is based on observations
done with the {\it XMM-Newton}, an ESA science mission with instruments 
and contributions directly funded by ESA Member States and the US (NASA). The 
Canadian Galactic Plane Survey is a Canadian project with international 
partners, and is supported by the Natural Sciences and Engineering Research 
Council (NSERC). NL and MS are thankful to the companies INSA and Selex I. S. resp. for
a financial support for the publication of this paper.
\end{acknowledgements}

\bibliographystyle{aa}
\bibliography{12253}
\IfFileExists{\jobname.bbl}{}
{\typeout{}
\typeout{****************************************************}
\typeout{****************************************************}
\typeout{** Please run "bibtex \jobname" to optain}
\typeout{** the bibliography and then re-run LaTeX}
\typeout{** twice to fix the references!}
\typeout{****************************************************}
\typeout{****************************************************}
\typeout{}
}

\end{document}